\documentclass[%
reprint,
superscriptaddress,
 amsmath,
 amssymb,
 aps,
prb,
]{revtex4-2}

\usepackage{graphicx}
\usepackage{bm}
\usepackage{hyperref}
\hypersetup{colorlinks=true,citecolor={blue},linkcolor={blue},urlcolor={blue}}
\usepackage{xcolor}
\usepackage{ulem}

\usepackage{soul}

\begin{document}

\preprint{APS/123-QED}

\title{QCL dynamics: thermal effects and rate equations beyond mean-field approach}


\author{Ivan I. Vrubel}
 \affiliation{Ioffe Institute, 194021 St.~Petersburg, Russia}
\email{ivanvrubel@ya.ru}

\author{Evgeniia D. Cherotchenko}
\affiliation{Ioffe Institute, 194021 St.~Petersburg, Russia}

\author{Georgii D. Miskovets}
\affiliation{Presidential Physics and Mathematics Lyceum 239,  St. Petersburg, Russia}

\author{Vladislav V. Dudelev}
\affiliation{Ioffe Institute, 194021 St.~Petersburg, Russia}

\author{Grigorii S. Sokolovskii}
\affiliation{Ioffe Institute, 194021 St.~Petersburg, Russia}





\begin{abstract}
The correct accounting for thermal effects is always a challenge when one needs to make quantitative predictions for any laser applications. In such complicated devices as quantum cascade lasers temperature strongly affects the operational conditions preventing reaching the CW mode as well as efficient lasing in pulsed regime.  Rate equations are the most effective and simple way to model laser dynamics. However, the conventional approaches operate under the mean-field approximation, considering finite number of population levels, generalizing the obtained results to the infinite number of cascades, and do not take heating into account. In this work we modify the conventional three-level rate equation approach by adding self-heating description and applying it to the calculation of QCL dynamics. As a result we show how temperature affects the threshold characteristics and build-up time and include electronic aspects to the description of QCL.
\end{abstract}

\maketitle


\section{Introduction} \label{sec:Intro}

The development of QCL-based technologies is on high demand due to applications arising in mid-IR range, like infrared counter-
measure technologies and optical communication developments, such as chaotic light
sources, spectroscopy, sensing, and many others. The very important, but less discovered part of QCL performance is the influence of thermal processes on the device operation\cite{sirtori02}. Considering the experiments, when the pump pulse has finite duration of the order of hundred nanoseconds, one can find that ohmic heating can change the initial lasing conditions. The investigation of temperature effects is obscure in both practical (experimental and technological) and theoretical studies and the development of effective heat dissipation (cooling) sub-
system and modelling of related thermal processes still attracts attention of researchers. The applied way has to solve the problem of effective thermal energy transfer from the active region to heat sink via media which does not distort electronic structure of cascades when they are coupled. At the same time the theoretical research suffers from poor fundamental description of thermal processes which take place in the active media under operation, therefore some phenomenological models\cite{sirtori02, hamadou08, agnew15, agnew16, yousefvand16, yousefvand17} are utilized for these investigations. An alternative way is a pure numerical simulation of the structure, that is successful, but needs a special software and gives a result for a specific design without perspectives for generalisation\cite{Lee2009}. Thus well advanced experiments suffer from the lack of theoretical predictions that are able to explain the observations not only qualitatively, but also give realistic estimates for further experiments.
{\color{black}This work aims at discovering the influence of heating on threshold characteristics of QCLs and their turn-on dynamics that need increased performance}.

Previously published theoretical studies\cite{hamadou09, hamadou13} report that the dependence of the turn on delay on pump current for QCL manifests behaviour contrasting the one for conventional for laser diodes\cite{coldren2012diode}. The last demonstrate the delay of the orders of nanoseconds due to the high rate of spontaneous emission\cite{Sokolovskii2012}.  However in QCL this rate is bypassed by the very efficient temperature dependent non-radiative scattering processes. Thus the theoretical predictions show the build-up times of the order of hundred picoseconds at the double threshold, which is the standard operational regime.
The recent experiment\cite{cherotchenko21} showed that the real value of the build-up time can reach the order of ten nanoseconds which completely contradicts the provided numerical estimations. In this work we want to investigate this discrepancy thoroughly by means of coupled rate equations.

The rate equations approximation is well-known and widely used\cite{hamadou09, cherotchenko21, cherotchenko16} mathematical approach allowing to describe system with classical or semi-classical behaviour.
The key advantages of this approach are simplicity and ability to probe basic laser characteristics with limited number of parameters.
This is possible due to usage of parameterized rates and cross-sections for relevant physical processes and consideration of the QCL in the spirit of mean field theory. The last means study of the complex QCL device consisting of a plenty of cascades by substitution with a single three level system, treated with a short set of differential equations.
In other words it is assumed that all individual processes for different cascades can be modelled by the single set of equations, while parameters for common phenomena (photonic mode) can be renormalized with respect to the number of cascades.

\begin{figure*}[ht]
\begin{eqnarray}
\begin{cases}
\begin{cases}
\frac{du_n}{dt} = S_n(t,T)-\left(\frac{1}{\tau_{32}(T)}+\frac{1}{\tau_{sp}}+\frac{1}{\tau_{31}(T)}\right)u_n-\Gamma c'\sigma {(u_n-m_n)} \frac{\Phi}{K V_c} \\ \\
\frac{dm_n}{dt} = \left(\frac{1}{\tau_{32}(T)}+\frac{1}{\tau_{sp}}\right)u_n-\frac{m_n}{\tau_{21}(T)}+\Gamma c'\sigma {(u_n-m_n)} \frac{\Phi}{K V_c} \\ \\
\frac{dl_n}{dt} = \frac{u_n}{\tau_{31}(T)}+\frac{m_n}{\tau_{21}(T)}-D_n(t,T)\
\end{cases}, \text{ for } n\in[1,K]\\ \\
\frac{d\Phi}{dt} =\Gamma c'\sigma \left(\sum\limits_{n=1}^K(u_n-m_n)\right) \frac{\Phi}{K} +\frac{\beta V_c}{\tau_{sp}} \sum\limits_{n=1}^Ku_n-(\frac{1}{\tau_{loss}}+\frac{1}{\tau_{out}})\Phi
\end{cases}
\label{sysrateeqns}
\end{eqnarray}
\end{figure*}

For mid-IR QCLs the standard rate equation system is discussed in \cite{hamadou09}. The simplicity of the equations for three levels features the possibility of analytical solution that coincides with the numerical simulations. At the same time this simplicity leads to some serious drawbacks. The first one was discussed in \cite{cherotchenko21}, where the same system was solved with pumping current profile close to the real experimental conditions. The experimental  behavior of the build-up time that is one of the main QCL dynamical characteristics contradicts with the theoretical predictions made in \cite{hamadou09, hamadou13}. The simplified analytic fails when the pumping differs from the step-like. The attempts to build the analytical model are discussed in \cite{kusakina22}, however the comparison with the experiment still remains insufficient. 

Closer look to the system of equations reveal the following reasons for that: the first one is irrelevant description of the lowest state, namely: the population of carriers in the lowest level does not participate in any physical process. Further analysis shows that the "injection" level can accommodate any number of electrons, which may lead to unrealistic equilibrium Coulomb charge of each cascade and  electrostatic field of ultimate strength. The next disadvantage is inability of the model to reproduce conditions for resonant tunneling of carriers through injector and related behavior of the V-I characteristic. Finally the temperature dependence of phonon - assisted rates is omitted, which ignores significant aspects of laser functioning\cite{evans07}. 

Some attempts to solve these problems were done in several theoretical works. In Ref.~\onlinecite{lops06} the thermal conductivity is assessed to develop optimal substrate configuration and geometry. More theoretical papers consider effect of thermal process through perspective of rate equations. On general scaling the main effect governing modification of QCL operation regime is the increase of the non-radiative relaxation rate\cite{sirtori02} with heating. Additionally more delicate macroscopic, quantum and phenomenological effects can be accounted for, e.g. electronic circuit-like behaviour of rate equations\cite{yong13}, effects of levels broadening on rate equations parameters\cite{hamadou08, wittmann08} or even results of Shrodinger-Poisson model solution\cite{agnew16, yousefvand17, kundu18}. Also in studies where a laser operates at non equilibrium regime the effect of active region self-heating process due to Joule heating \cite{agnew16, kundu18} is considered.

\section{Model}

In this work we are focused on the build-up time, thus we need to compile the consistent model, which on the one hand adopts the most important features listed in the previous paragraph, while remains solvable with the use of only fundamental constants and basic physical principles on the other. In our description we use the following estimations:
\begin{itemize}
    \item In order to avoid the use of artificial approaches to account for the electronic circuit behavior we model all cascades implicitly, the power supply is considered as a current source, thus the voltage applied to the system is governed by the microscopic processes of the system;
    \item each cascade contains three levels - injection level and two lasing levels;
    \item threshold current depends on temperature exponentially in the vicinity of room temperature (RT); we neglect potential small variations\cite{evans07} of the dependence.
  
\end{itemize}

\subsection{Rate equations approach}

Here we introduce the modified system of rate equations, that satisfies the following conditions:
\begin{itemize}
\item  accounts for the thermal processes
\item satisfies Kirchhoff's circuit laws to conserve system electrical neutrality
\item makes the lower level of the system accountable for modelling
\end{itemize}

\begin{table}
\caption{Parameters}
\begin{tabular}{c c }
parameters & values\\ \hline \hline
$\tau_{31}^{RT}$, $\tau_{32}^{RT}$ and $\tau_{21}^{RT}$  & 1.5~ps, 0.8~ps, 0.7~ps \\ 
$K$ & 50 \\ 
$c'$ &  10$^{10}$~cm~s$^{-1}$ \\ 
$\sigma$ & 10$^{-13}$~cm$^2$ \\ 
$\beta$ & 3~10$^{-3}$ \\ 
$\Gamma$ & 0.5 \\
$\tau_{sp}$ &  1~$\mu$s   \\ 
$l$ & 0.5~cm \\
$R_1$=$R_2$ &  0.3 \\ 
$L_c$ & 0.3~cm \\
$S_c$ & 6~10$^{-4}$~cm$^2$ \\
$V_c$ &  3~10$^{-9}$~cm$^3$ \\ 
$n_0$ &  10$^{17}$~cm$^{-3}$\\
$U_b$ &  0.3~V \\ 
$\tau_{inj}$  &  1~ps \\ 
$C_c$  & 0.3~J~(g~K)$^{-1}$ \\
$T_{imp}$  & 100~ns \\
$T_{rise}$  & 10~ns \\
$n_0$ &  10$^{17}$~cm$^{-3}$ \\
$\hbar\omega_0$ & 30~meV \\
$\rho$  &  5~g~cm$^{-3}$ \\
$\epsilon$ & 10 \\
{\color{black} $h\nu$} &{\color{black} 0.15~eV} \\
\end{tabular}
\label{tbl:syseqparams}
\end{table}

If one attempts to upgrade the conventional 3-level rate equation model to fix the drawbacks highlighted in Section~\ref{sec:Intro}, one would have the system of rate equations as in eqs.~(\ref{sysrateeqns}). Here $u_n$, $m_n$, $l_n$ are time dependent concentrations [cm$^{-3}$] of electrons in upper, middle and lower levels occupying cascade number ``n'', where lasing transition occurs between the upper and middle level, $\Phi$ is a time dependent number of photons in the active region (dimensionless), $\tau_{31}(T)$, $\tau_{32}(T)$ and $\tau_{21}(T)$ -- temperature dependent phonon-assisted non-radiative relaxation lifetimes [s],  $K$ -- a number of cascades, $c'$ -- a speed of light in the media [cm~s$^{-1}$], $\sigma$ -- a stimulated emission cross-section [cm$^2$], $\beta$ -- efficiency of spontaneously emitted photon capturing into the waveguide mode (dimensionless), $\Gamma$ -- mode confinement factor (dimensionless), $\tau_{sp}$ -- spontaneous radiative transition (u$\rightarrow$m) lifetime [s], $\tau_{loss}$  and  $\tau_{out}$  -- lifetimes of photons loss due to inevitable irreversible losses inside the media and output loss correspondingly [s], $S_n(t,T)$ and $D_n(t,T)$  are source and drain for electrons in cascade ``n'' in the following form: 

\begin{eqnarray}
\begin{cases}
S_1(t,T) \equiv D_K(t,T) \equiv \frac{I(t)}{V_c} \\
\text{otherwise}\\
S_n(t,T) = D_{n-1}(t,T) = \frac{l_{n-1}(t)}{\tau_{inj}}exp\left(-\frac{e(U_b-U_{n-1}(t))}{k_BT}\right)
\end{cases}
\end{eqnarray}

\noindent where $I(t)$ -- power supply current profile [A], $k_B$ -- Boltzmann constant [eV~K$^{-1}$], $V_c$ -- cascade volume [cm$^3$], $T$ -- local temperature of the active region [K], $U_b$ -- nominal voltage bias between cascades [V] allowing resonant tunneling current through injector at temperature of 0~K, $U_n(t)$ -- time dependent voltage between ``n'' and ``n+1'' cascades [V], $\tau_{inj}$ is a lifetime of an electron on the lower level at $U=U_b$ [s]. The exponential part is responsible for the carrier capability to tunnel from the one cascade to another. Initially we assume that the system of cascades is electrically neutral with doping level of $n_0$ [cm$^{-3}$].

For calculation of the bias voltage between cascades we use the superposition principle. We assume that charge of each individual cascade creates electrostatic field with specific strength. Thus the total field strength between ``n'' and ``n+1'' cascades may be described as:

\begin{multline}
    E_n(t)=\frac{V_c}{2 \epsilon \epsilon_0 S_c} \sum\limits_{i=1}^{K} (-1)^p  \left( u_i(t)+m_i(t)+l_i(t)-n_0 \right),\\
    \text{ where } \begin{cases} p=0, \text{ if } i\le n \\ p=1  \text{ if } i > n  \end{cases}
\end{multline}

\noindent where $S_c$ -- a cascade area [cm$^2$], $\epsilon$ -- a permittivity of a material. 

The rate of photons loss due to absorption in the cavity obviously equals:

\begin{equation}
\tau_{loss}=\frac{l}{c'},
\end{equation}

\noindent where $l$ is a mean free path of the photon in the cavity [cm]. To assess the rate of photon leakage through the facet out of the laser media we use the standard formula\cite{hamadou09}:

\begin{equation}
\tau_{out}=-\frac{2 L_c}{c'
~\text{ln}(R_1R_2)}, 
\end{equation}

\noindent which results in the power of emitted radiation in the following form:

\begin{equation}
P(t)= \frac{h\nu\cdot \Phi}{\tau_{out}}.
\label{optoutpow}
\end{equation}

The last option of the proposed model is the temperature. Since we model a QCL working in pulse regime the heat released during the pump can be considered as absorbed by the cascade material. Assuming linear behaviour of the thermal capacity and presence of the heat sink having stable temperature and playing the role of a reservoir, one may deduce the local temperature of the cascade as:

\begin{multline}
T(t)=T_{hs}+\int\frac{dQ}{C_c\rho V_c}=T_{hs}+\int\limits_0^{t}\frac{I(x) U_{n}(x) dx}{C_c \rho V_c}\approx \\ \approx T_{hs}+\frac{I U_{n} t}{C_c \rho V_c}    
\label{JouleHeat}
\end{multline}

\noindent where $T_{hs}$ is a temperature of a heat sink, $C_c$ and $\rho$ -- are cascade material heat capacity [J~(g~K)$^{-1}$] and density [g~cm$^{-3}$].  Simple estimations performed with the values presented in Table~\ref{tbl:syseqparams} and typical experimental conditions, when a current pulse of 1~A is applied for 100~ns, provide the excess temperature of an active region up to 10~K.

Since we include temperature in the model we cannot neglect a temperature dependence of any phonon-assisted relaxation rates. We use the following estimation of the rates:

\begin{equation}
\tau (t) =  \tau^{RT} e^{-\frac{k(T(t)-RT)}{\hbar\omega_0}}
\label{phratetempdep}
\end{equation}

\noindent where $\tau^{RT}$ is a value of a rate at room temperature ($RT$), $\hbar\omega_0$ is an energy of the effective phonon mode. The value of $\hbar\omega_0$ is fitting parameter, which is adjusted to reproduce the temperature behaviour of the threshold current. 
To model pump pulse with non-zero rise time we used the following approximation:

\begin{equation}
I(t)= I_0(1-e^{-\frac{t}{T_{rise}}}),
\label{currentpulse}
\end{equation}

\noindent where $T_{rise}$ -- is a rise time (in fact RC constant of a pump source) of a current pulse [s]. $I_0$ -- is an equilibrium value of pump current [A].

\subsection{Modelling}

\begin{figure}[]
\includegraphics[scale=0.31]{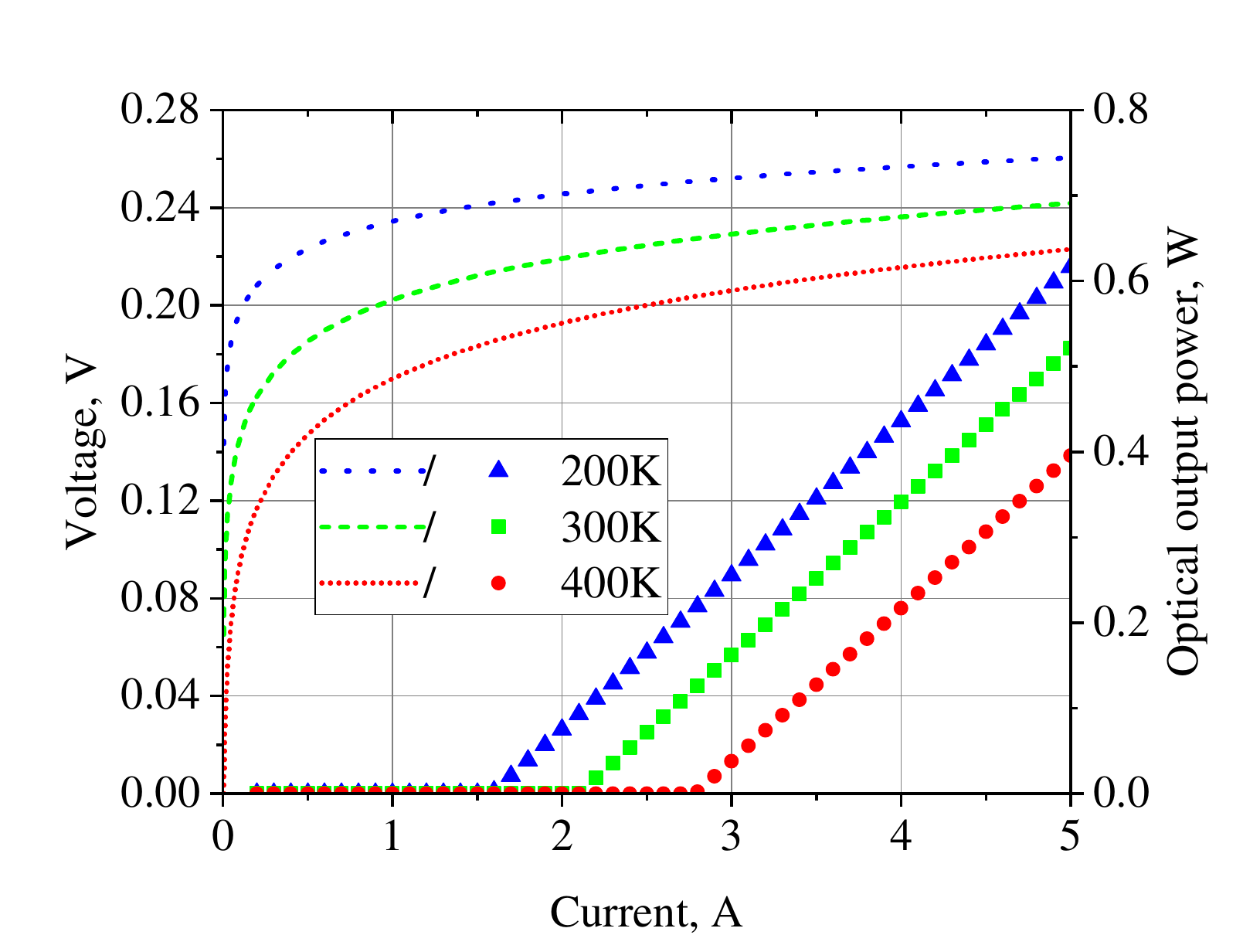}
\caption{{\color{black}V-I (per one cascade) and L-I characteristic calculated with the proposed model, simulating operation of a QCL. The optical output power is the value proportional to the maximum photonic mode population (eq.~(\ref{optoutpow})). 
The temperature dependence is due to variable rate of non-radiative (phonon-assisted) relaxation, which is a function of temperature-dependent (see eq.~(\ref{phratetempdep})) average phonon occupation number.
The pump of the model QCL is a current source, analogous to a FET-based power supply. The voltage is the observable value of the inter-cascade potential in the end of a pump pulse for a given current. One can find that the model manifests typical p-n -like behaviour. The temperature dependence of V-I characteristic is due to implicit modelling of the inter-cascade current relaxation rates.}}
\label{fig:cctemperature}
\end{figure}

\begin{figure}
\includegraphics[scale=0.31]{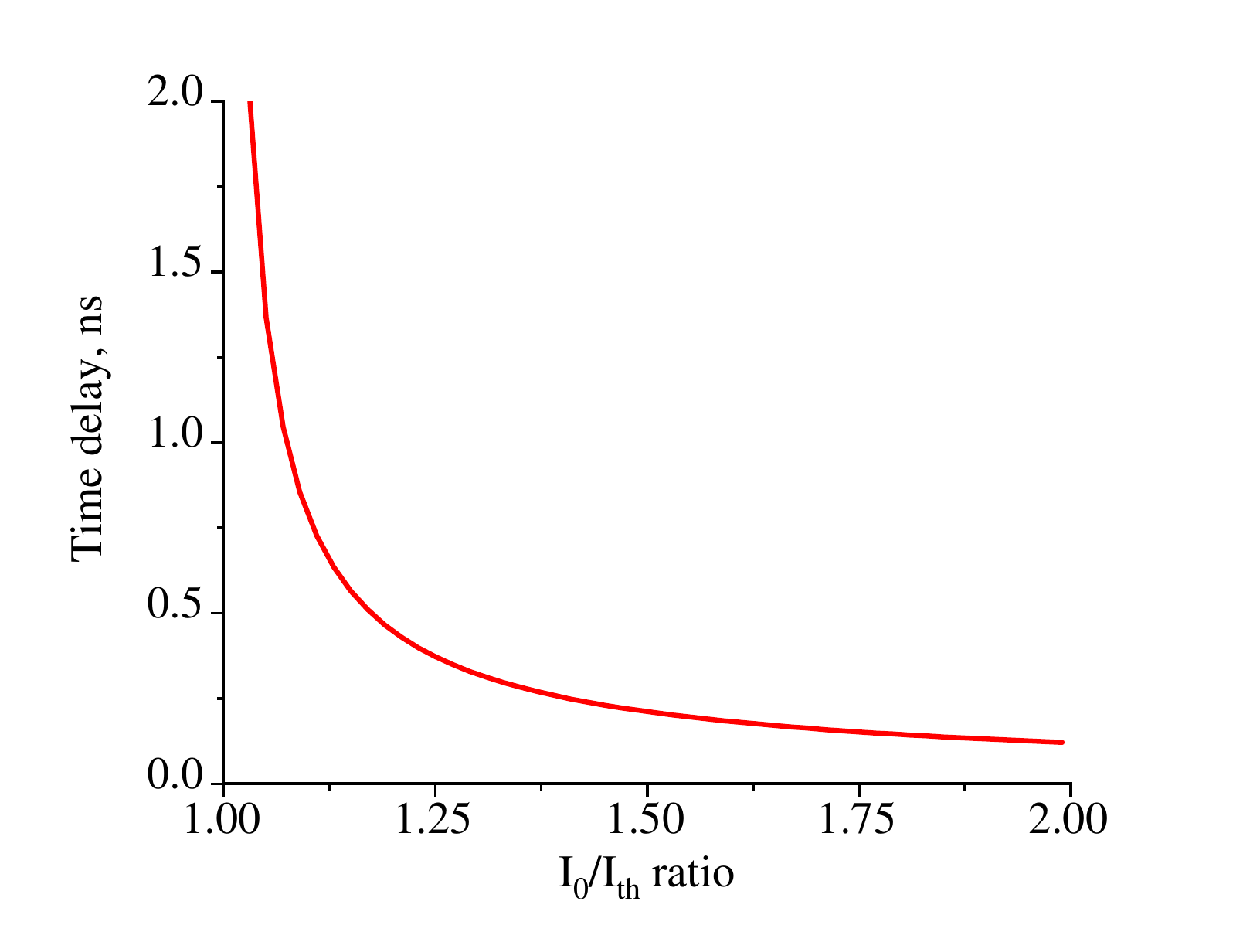}
\caption{{\color{black}10\% Build-up time plotted within the proposed approach  with no self-heating processes taken into account. The threshold is calculated using light-current characteristic, based on the peak values of lasing emission.}}
\label{fig:delayhside}
\end{figure}

In our work we mainly focus on the effects manifesting when an active region local temperature rapidly grows with respect to the heat sink. However, proper model parameterization  adequately reproduces more general effects. The most illustrative of them is a dependence of L-I and V-I characteristic on significant variations of the heat sink temperature, when changes of the local temperature are negligible. Interestingly, the later function is not common for such kind of modeling and available due to explicit evaluation of each system cascade, that additionally validate the proposed temperature dependence.
Fig.~\ref{fig:cctemperature} demonstrates  I-V curves as a function of heat sink temperature.  The parameters used in the calculation are provided in Table~\ref{tbl:syseqparams}. 
L-I curves, shown in Fig.~\ref{fig:cctemperature}, plotted as a function of heat-sink temperature demonstrate similar behavior as in \cite{evans07}. To verify the model we applied the Heaviside-type step-like pumping in order to reproduce the build-up time within this approach; The threshold current is defined from the L-I characteristics and equals to $I_{th}$=2.1~A at room temperature (RT). Fig.~\ref{fig:delayhside} reproduces the asymptotic behavior of the build-up time similar to that shown in Ref.~\onlinecite{hamadou09}.

{\it Effect of self-heating}:
\begin{figure}
\includegraphics[scale=0.31]{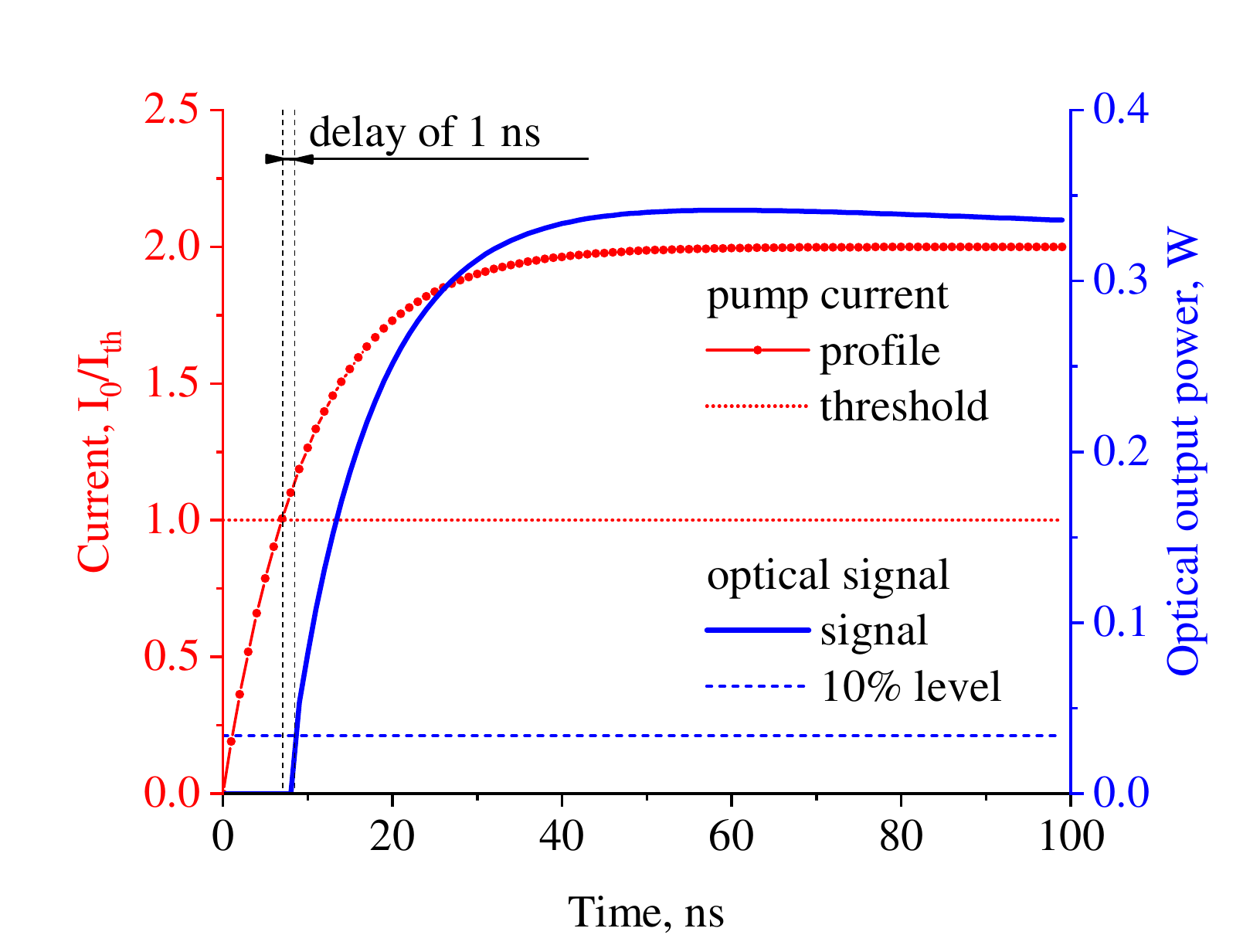}
\caption{Modelled time dependent waveforms of optical signal and corresponding pump current of doubled threshold}
\label{fig:doubthresh}
\end{figure}

\begin{figure}
\includegraphics[scale=0.31]{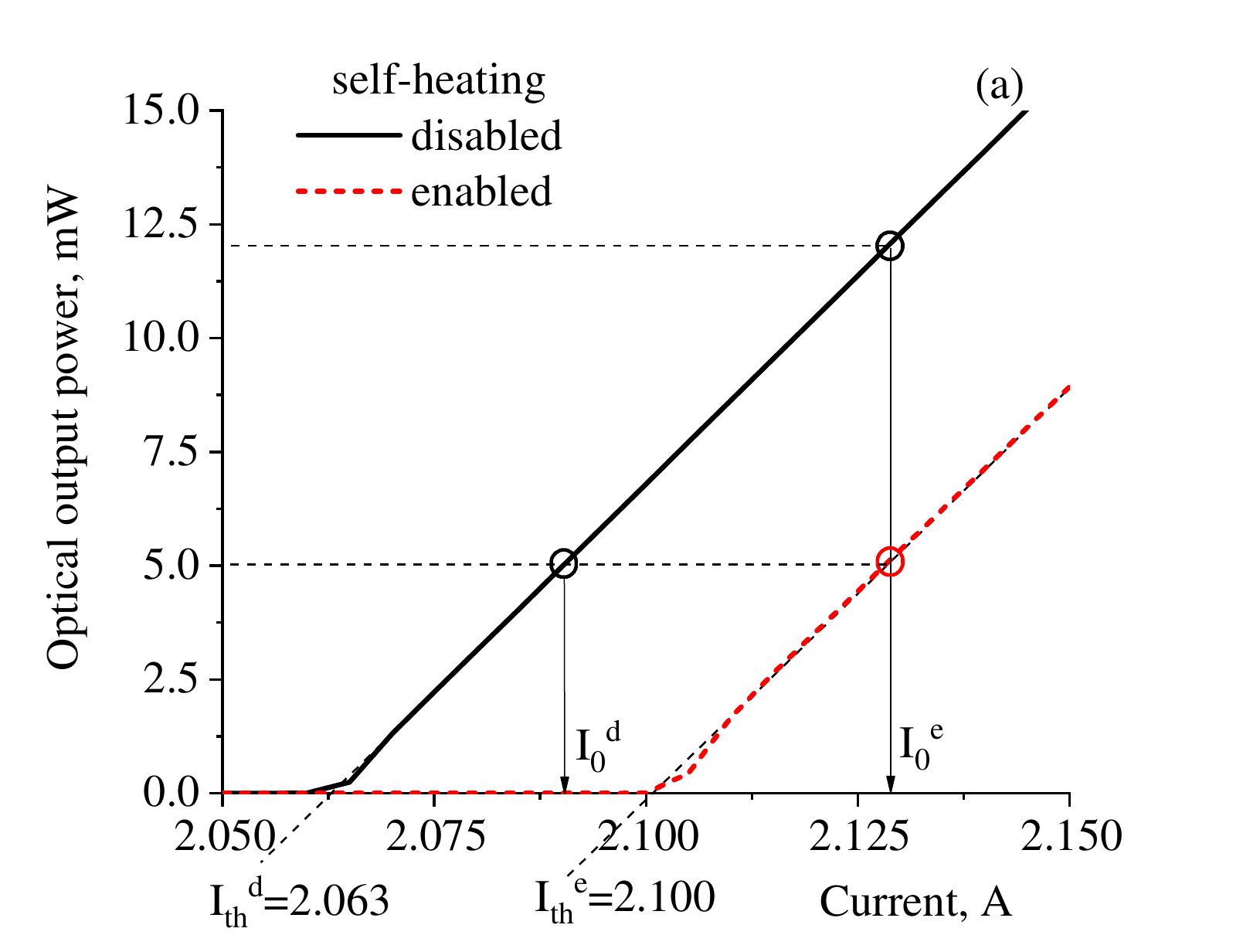}
\includegraphics[scale=0.31]{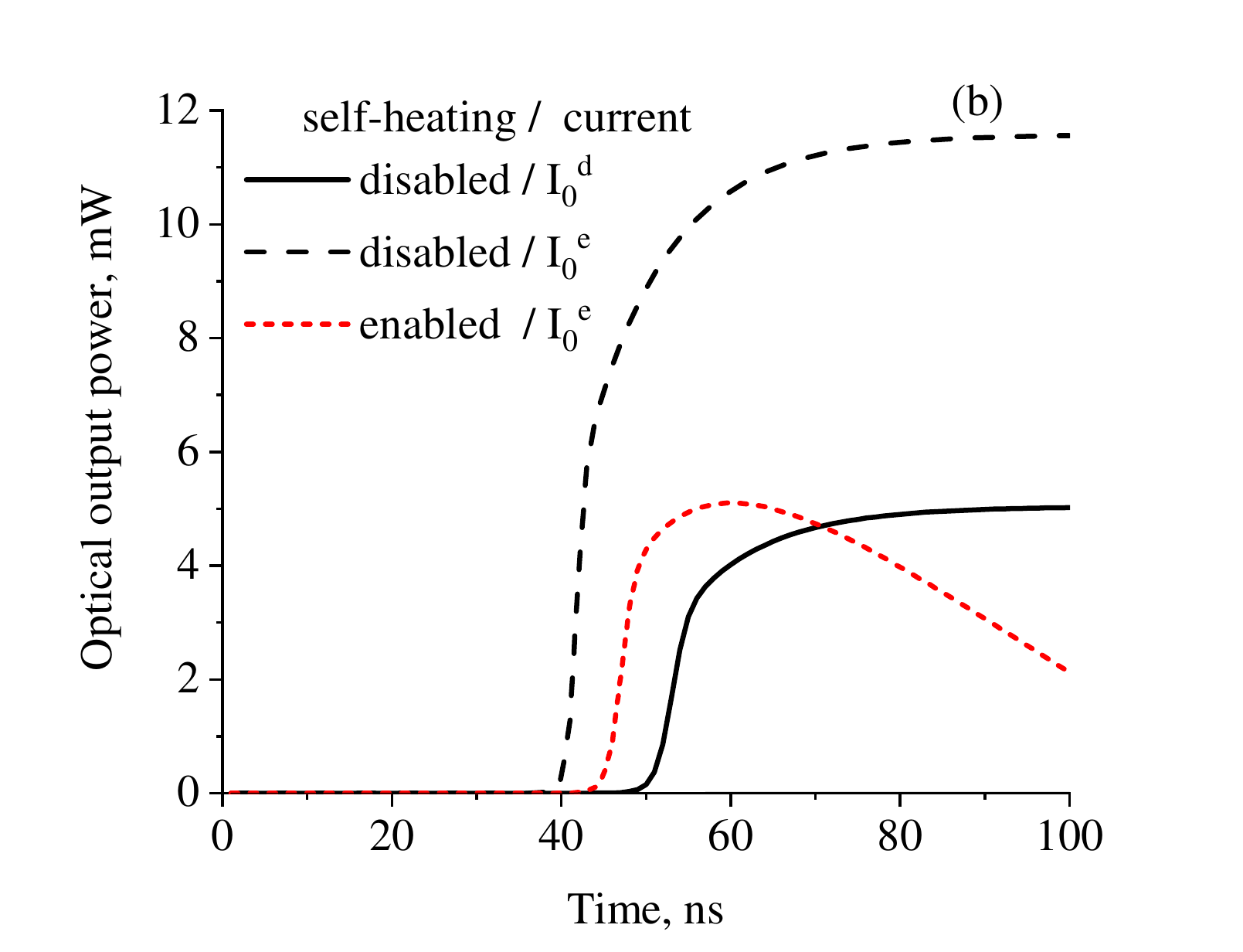}
\caption{{\color{black}(a)~Near threshold zoom of light-current characteristic modelled at room temperature ($T_{hs}$=300~K). The two curves differ only in the status of the self-heating option. In the curve with self-heating option enabled higher level of generation threshold manifests due to increasing phonon-assisted relaxation rates stimulated by the instantaneous local heating during pump pulse.
(b)~Modelled time dependent optical signals calculated for pump currents marked in panel (a) as I$_0$$^\text{d}$ and I$_0$$^\text{e}$. Black solid line indicates the optical signal  under pulsed pumping with I$_{max}$= I$_0$$^\text{d}$. Black dashed line indicates the temporal evolution of the lasing emission for the pump current I$_0$$^\text{e}$ with disabled heating, red dashed line - with enabled one. Non-monotonous behaviour of the optical signal modelled with enabled self-heating results from the increasing predominance of the phonon-assisted processes during pump pulse with the current of I$_0$$^e$} 
}
\label{ZOOMthresh}
\end{figure}

Enabling self-heating during pump pulse by taking into account the Joule heating in eq.~(\ref{JouleHeat}) and keeping all parameters of the model one can get that the threshold current grows up. 
Obviously this is due to the heat produced before beginning of generation, that warms a sample reducing the probability of stimulated emission compared to the non-radiative processes. 
Experimentally this is realized when the pumping is different from the step-like function and shows the non-zero rise time as done in \cite{cherotchenko21} and shown in Fig.~\ref{fig:doubthresh}.
To make a closer look and comparison in Fig.~\ref{ZOOMthresh}(a) we plot two L-I characteristics where the red-one includes the Joule heating released in the active region. 
Obviously the longer the pumping rise time is, the higher threshold current will be technically detected. 
Thus we observe the dynamical threshold variation and come to a problem of correct threshold determination during the experiment.
Apart from that one can see that due to the Ohmic processes the time dependence of optical signal is non-monotonous in comparison to the case without heating (see the red and black dashed curves in Fig.~\ref{ZOOMthresh}(b)). This correlates with the experimental data shown in \cite{cherotchenko21} and can be explained by the thermal processes distorting delicate balance between probabilities of non-radiative relaxation and stimulated emission.

\begin{figure}
\includegraphics[scale=0.31]{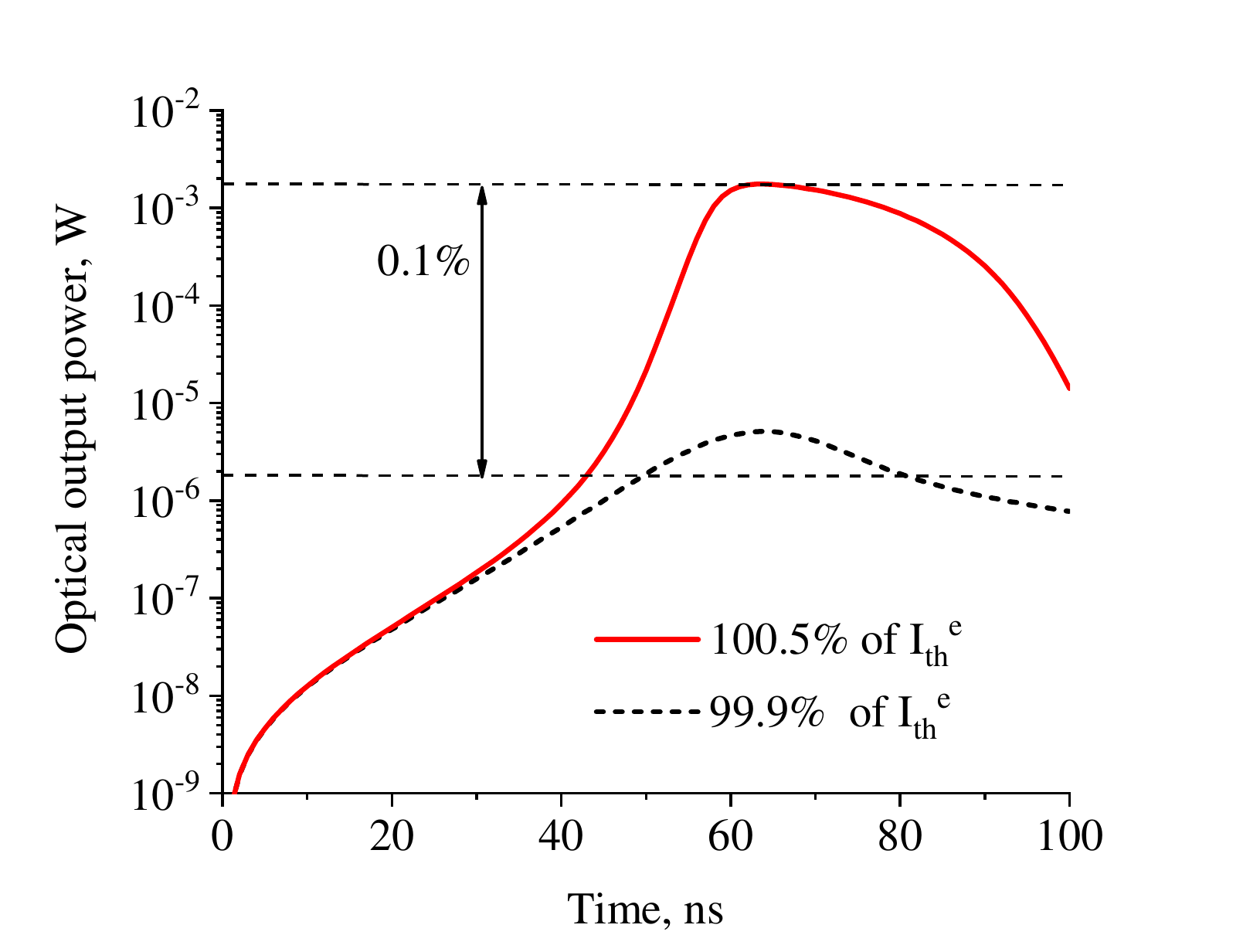}
\caption{Modelled optical signal of laser operating at 0.999 and 1.005 of threshold current. When a current pulse of 100.5\% of threshold is applied the value of laser emission is 2~mW (see also Fig.~\ref{ZOOMthresh}(a)). Thus the value of 0.1\% of peak intensity (2~$\mu$W) used to determine beginning of lasing ambiguously merges with the spontaneous emission, which makes plotting of build up dependence impossible for current values approaching to the threshold too close.}
\label{fig:lowestcurrent}
\end{figure}

\begin{figure}
\includegraphics[scale=0.31]{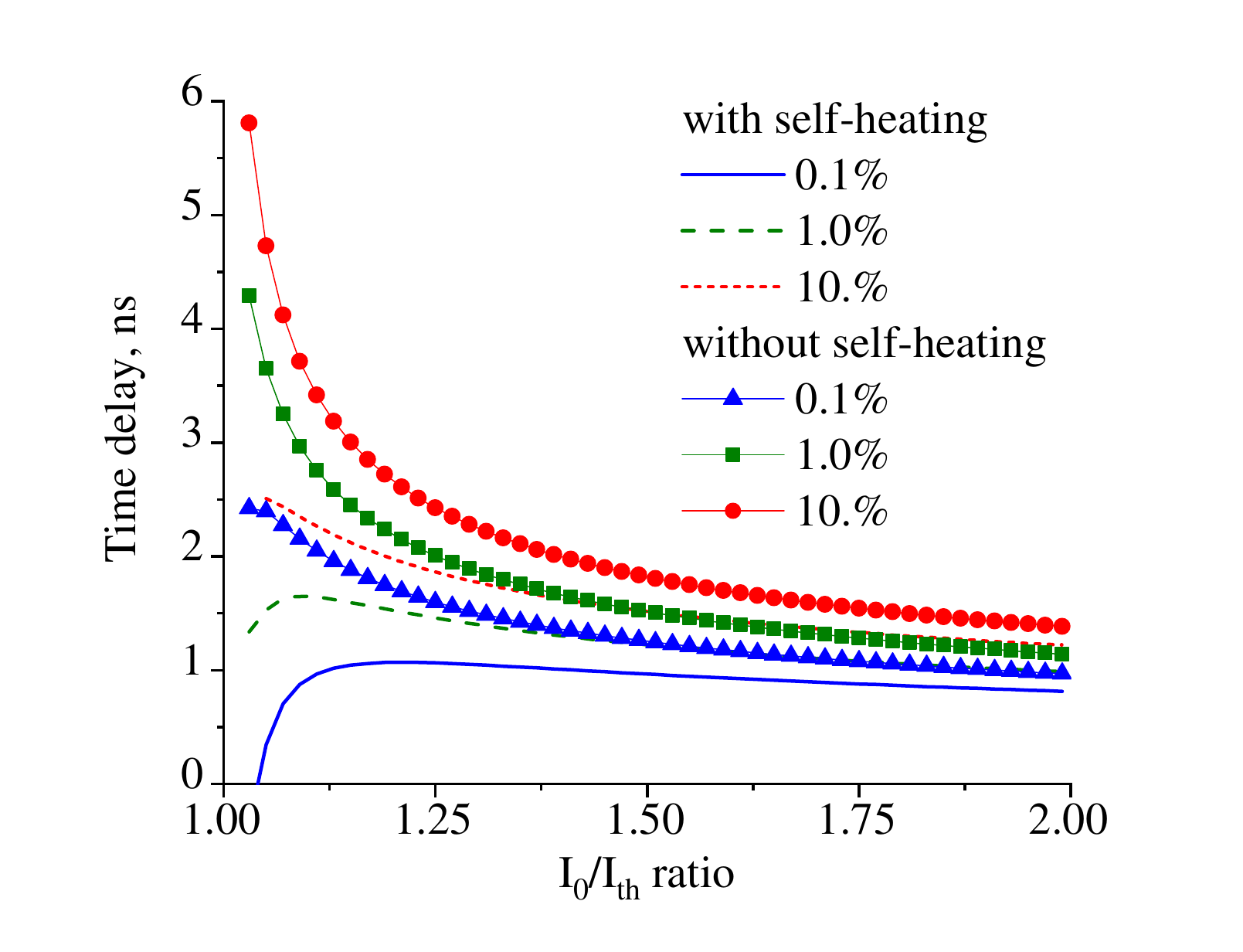}
\caption{{\color{black}Time delay between reaching of a threshold current by a pump pulse (where threshold is calculated using L-I characteristic plotted with the peak values of light emission) and beginning of a laser radiation emission. The low level to determine ``presence'' of a radiation for an optical signal is set to 0.1\%, 1.0\% and 10\% of a maximum instantaneous signal power.}}
\label{fig:realdelay}
\end{figure}

{\it Build-up time estimations}:
Here we discuss the self-heating impact on the build-up time, which is important parameter for such applications as LIDAR systems and telecommunication technologies. Practically it can be defined as the period between reaching threshold current and beginning of laser radiation emission (usually 10\% of maximum intensity is considered). As it follows from the previous results the generation threshold is ambiguous, significantly depends on both intrinsic properties and specific experiment conditions and cannot be considered as a fundamental constant of a QCL even at a stable heat sink temperature. Therefore for measurements including the threshold as a main parameter all details of the experimental protocol must be accounted for.
In \cite{cherotchenko21} it was shown that even under step-like pumping the build-up time demonstrates the peak in the near-threshold (of the order $I_0\approx$~1.001$I_{th}$) region of the pumping currents. 
On the one hand this peak directly follows from the analytical expression  for the build-up time given in \cite{hamadou09}. On the other hand the time delay estimation has to be performed with caution, because trigger levels equal to 0.1$\%$, 1$\%$ and 10$\%$ of maximum (equilibrium) lasing intensity must be formed by stimulated emission rather that spontaneous one. 
{\color{black}The last should be checked each time with the L-I curve. Fig.\ref{fig:lowestcurrent} illustrates this idea: the two waveforms of a laser operating at 0.999$I_{th}$  and 1.005$I_{th}$ are shown together with 0.1\% level intensity for the later}. 
One can find that the last merges with the spontaneous emission region, making the definition of the build-up time uncertain. Fig.\ref{fig:realdelay} shows the numerical simulation of the build-up time, considering the self-heating and absence of self-heating processes. It is clearly seen that including of self-heating shifts the peak value of a build-up time to higher pumping currents, that correlates with experimental data in Ref.~\onlinecite{cherotchenko21}. 

\section{Discussion }

The proposed model includes all the physical process responsible for accurate modeling of QCL build-up time. It naturally includes and combines microscopic description of quantum processes and macroscopic description of QCL as an electronic device.
This advantage is brought about the use of explicit consideration of each cascade in the active region and relevant modelling of the injector and pump current. Together with the introduced inter-cascade current temperature dependence this allows simulation of the V-I characteristic, which cannot be presented as an observable of modelling in the mean-field approximation. On the one hand results obtained in the Poisson-Schrodinger approach require introduction of parameter of an external electrostatic field. On the other hand all models evaluating cascades in the mean field manner are not able to reproduce any fields in the structure, because their lowest level is irrelevant even if it is modelled. Our model behaves adequately in the sense of both concepts, in particular the structure manifests uniform electrostatic field, which strongly depends on pump current and temperature. The electrostatic field emerges when power supply pumps the two outer cascades by positive and negative total charges. Interestingly, the rest of cascades stays neutral in any operation regime. Our calculation shows that charging of outer cascades occurs prior to reaching the threshold current, thus it does not affect significantly the build-up time (see Appendix~\ref{app:electronics}). Obviously, the manifesting electronic-like effects can be utilized as a basis for further upgrade of the model with the Poisson-Schrodinger self-consistent loop.

\subsection{Time delay as a specification parameter of QCL}

The thorough modeling of the QCL operation demonstrate that the build-up time is not a fundamental characteristic of a device. In Fig.~\ref{fig:realdelay} one can find zero delay at a current above the threshold current. The reason for that is as follows.
The common experimental method to calculate the threshold in pulse pump regime is based on the L-I characteristic processing. Usually the value of the threshold current is calculated as a root of the linear function fitted in the lasing domain of L-I curve. However if the thermal energy release in the active region of QCL is adiabatic compared to heat dissipation over TEC, then the L-I characteristic is distorted by Ohmic heat in particular near the threshold. In fact depending on the pump pulse shape and its equilibrium value the heat released in the media may vary significantly. 

\subsection{Quantum noise and build-up jitter} 
The quantitative description of behavior of a vast ensemble of quantum particles allows the use of semi-classical approach such as rate equations, however such approaches provide only average values of populations. Thus we are able to assess only the mean value of build-up time by means of Fig~\ref{fig:realdelay}. Nevertheless some applications especially LIDAR and alike require not only the mean-value but also its uncertainty (time jitter) to meet strict requirements. Let us assess the possible fluctuations for the moment of laser emission beginning. If an average number of photons is low then the emission is randomly triggered by the first photon emitted in the cavity mode, which creates an uncertainty. Otherwise, laser have photons in the cavity to gain their population immediately. In this case the jitter is governed by the lifetime of photons in the cavity. Our results show that each QCL cascade has about 10$^{7}$~electrons on the upper level under pump current of about threshold value, thus the rate of photonic mode filling due to spontaneous emission is about:

\begin{equation}
    \beta K \frac{u_n V_c }{ \tau_{sp}}\approx 1.5\cdot 10^{12}~\text{s}^{-1},
\end{equation}

\noindent while the lifetime of a photon in the cavity is about 7~ps. Multiplying these values one obtains that the average number of photons  occupying the cavity mode prior to beginning of lasing is about:

\begin{equation}
    \overline{\Phi}=10.
\end{equation}

Here the calculated value suggests that working regime of the modelled QCL is far from the condition, where beginning of the optical signal generation strongly depends on the moment of emission of ``the first'' spontaneous photon in the cavity mode. Practically that conclusion allows to assume jitter to be in the order of photon lifetime in the cavity.

\subsection{Heat dissipation}

Let us consider general estimations on the thermal regime of average laser operation neglecting planar effects and asymmetry of a typical QCL in $z$ direction. Assume the QCL lasing structure is placed in the environment (substrate and cladding layers) made of material with high thermal conductivity e.g. InP. Applying the current pulse with non-zero rise time one warms up the active region of the laser by the Joule heat. Our estimations show that with a pulse length of the order of 100~ns\cite{Dudelev2021} the temperature may increase by units of degrees. Now consider the evolution of the generated energy by means of heat equation:

\begin{eqnarray}
    \frac{\partial T}{\partial t} = \alpha \nabla^{2} T
    \label{heatequation}
\end{eqnarray}

\noindent where $\alpha$ is a thermal diffusivity for the substrate/cladding [cm$^2$s$^{-1}$]. The general solution for such a problem is a convolution of the initial temperature distribution along $z$ axis with the fundamental solution. Nevertheless in our case assuming that the heated layer is thin enough compared to the cladding thickness we can use it directly for general estimations:

\begin{eqnarray}
    T(z,t)=\frac{1}{\sqrt{4 \pi \alpha t}} e^{-\frac{z^2}{4 \alpha t}}.
    \label{thermcapfundsol}
\end{eqnarray}

In eq.~(\ref{thermcapfundsol}) the term $\alpha t$ governs the FWHM-like ($\sigma$ for Gaussian distribution) parameter for spatial distribution of a heat ($\delta z_{heat}$).
Consider the effective thickness of a heated layer at power off condition and 10~ns, 100~ns and 1~$\mu$s after, i.e. the specific shape of the fundamental solution at 100~ns, 110~ns, 200~ns, and 1100~ns.

The parameter of thermal diffusivity depends strongly on the material of the substrate, its doping, temperature,  etc., thus in the case of InP claddings we get the upper limit (highest possible value) of 1~W(cm~K$^{-1}$ \cite{glazov77}.
Other parameters required to calculate $\alpha t$ are presented in Table~\ref{tbl:syseqparams}. Combining all together one can get the formula for the rate of heat spreading over device:

\begin{eqnarray}
    \delta z_{heat}=2.4\sqrt{2 \alpha t} = 2.4 \sqrt{ \frac{2\kappa t}{C_c \rho}}
\end{eqnarray}

Implementing the mentioned time intervals one obtains approximately 8~um, 9~um, 12~um and 30~um. Such numbers suggests that for lasers working in pulse regime the effect of delay time modulation by self-heating should manifest for almost all III-V based structures, because the heat produced during approaching of the current to the threshold value (dozens of nanoseconds) indeed can be considered as adiabatic compared to heat dissipation.
\section{Conclusion}
In this work we solve the system of rate equations, that explicitly considers processes in all cascades of QCL and takes into account the adiabatic self-heating in the system. As a result we are able to obtain reliable V-I characteristic, show the strong temperature dependence of the threshold current and determine the build-up time that is strongly related to the latter. The proposed model reproduces the temporal behavior of the lasing emission during the pump pulse, explains the recent experimental data, presented in \cite{cherotchenko21} and proves that temperature control is extremely important for the experimental determination of the build-up time in QCL.

\begin{acknowledgements}
I.V., E.C., V.D., and G.S. acknowledge the support from the Russian Science Foundation (project 21-72-30020)
\end{acknowledgements}

\appendix

\section{Electronics} \label{app:electronics}
The important feature of the proposed model is an ability to capture details of QCL electrical functioning. In fact, total Coulomb charges of all inner cascades were equal to zero for all performed calculations. While the total charges of external ones had opposite signs and equal values. The value of this sign can be assessed via simple model of parallel-plate capacitor, where capacity reads as follows:

\begin{equation}
    Cap=\frac{\epsilon \epsilon_0 S_c}{d} \approx 3\text{pF}.
\end{equation}

The power supply provides pump pulse obeying linear dependence in the very beginning:

\begin{equation}
    I(t)\approx I_0 \frac{t}{\tau_{rise}}.
\end{equation}

The applied voltage governs the charge flow through the system. Assuming that the value of potential aligning cascades to operating regime is about $U_{on}$=10~V one can estimate the time needed for power supply to enable the current throgh capasitor as:

\begin{equation}
    U(t)=\frac{\int_0^t I(x)dx}{Cap} \rightarrow t_{on}=\sqrt{\frac{2 \tau_{rise} Cap  U_{on} }{I_0}} \approx 1~\text{ns}
\end{equation}

As it can be clearly seen in Fig.~\ref{fig:doubthresh} the time required to reach threshold current is about 10~ns, while the estimation on time needed for power supply to charge parasitic capacitance of QCL to working voltage is about 1~ns.

\bibliography{citations}

\end{document}